\documentclass[preprint,aps,amsmath,amssymb]{revtex4}
\usepackage{graphicx}
\usepackage{dcolumn}
\usepackage{bm}
\usepackage[bookmarks,hypertex]{hyperref}

\begin{document}

\title{First-Principles Study of Elasticity and Electronic Structure of Incompressible Osmium Diboride}

\author{Z. F. Hou}
\email{zfhou@fudan.edu.cn} \affiliation{Department of Physics,
Fudan University, Shanghai, P. R. China 200433}
\date{\today}

\begin{abstract}
Recently, osmium diboride (OsB$_2$) has attracted considerable
attention as an incompressible and hard material. We
investigate the structural property, elastic constant, and
electronic structure of orthorhombic OsB$_2$ by the
first-principles total energy calculations. The calculations are
performed within the density functional framework using the
projector augmented plane wave method. The structural property and
bulk modulus of OsB$_2$ compare well with experimental data. The
nine independent elastic constants of orthorhombic OsB$_2$ at
zero-pressure have also been calculated by symmetry-general
least-squares extraction method.  We have analyzed the mechanical
stability of orthorhombic OsB$_2$ in term of the calculated elastic
constants. A detailed study of the electronic structure and the
charge-density redistribution reveals the features of strong covalent
B-B and Os-B bondings in orthorhombic OsB$_2$. The orbital
hybridization and the characteristics of bonding orbitals in OsB$_2$
are identified. Orthorhombic OsB$_2$ exhibits a metallic character
and the states at Fermi level mainly come from the $d$ orbital of Os
atoms.

\end{abstract}
 \maketitle
\section{\label{sec:intro}Introduction}
To design and explore ultrahard materials is of great scientific and
practical interest in both experimental and theoretical
studies\cite{Veprek99,Brazhkin02,Haines01,Liu89,Jhi99,Kaner05},
because the class of hard materials has a wide variety of important
technological applications ranged from cutting and polishing tools
to wear-resistant coating \cite{Solozhenko01,Jhi99,Brazhkin02}.
Stiffness reflects the elastic compressibility of materials, which
is related to the bulk modulus, shear modulus, and the Young's modulus.
Hardness  describes the resistance of materials to plastic
deformation \cite{Kaner05}. It is recognized that there is a good
correlation between shear modulus and hardness
\cite{Brazhkin02,Haines01,Teter98} and a three-dimensional network
composed of short , strong bonds is critical for hardness of
material \cite{Kaner05}.  Many  synthetic and theoretical efforts
are made to design new ultrahard materials in past decades and two
main approaches have been summarized \cite{Kaner05} as the
following: a) combination of  light elements (such as boron,
nitrogen, and/or oxygen) to form short covalent bonds and b)
compounding of elements with very high densities of valence
electrons to ensure that the materials resist being squeezed
together. Cubic BC$_2$N, an example of the first approach, was
synthesized as a dense ternary phase in the B$-$C$-$N system
\cite{Solozhenko01} and found to be an extreme hardness second only
to diamond \cite{Solozhenko01,Zhang04}.   For the second approach,
one usually uses transition metals having a high bulk modulus but
low hardness to combine with small, covalent bond-forming atoms
\cite{Kaner05}. For example, highly incompressible phase of RuO$_2$
\cite{Haines93} owes to a strong covalent bonding between  ruthenium
$d$ states and oxygen $p$ states \cite{Lundin98}.

Other representative example for the second approach is the
compounds related osmium (Os) element. The pure Os metal was
measured to have a high bulk modulus of 462 GPa \cite{Cynn99},
411$\pm$6 GPa \cite{Occelli04}, and 395$\pm$15GPa \cite{Takemura04}
by different experimental methods, but found to have a  low hardness
of only 400 kg/mm$^2$ \cite{Shackleford01}.  These features of could
be understand that  \cite{Kaner05}   the high bulk modulus of Os
mainly owes to the high valence electron density (0.572 $e$/\AA$^3$)
and the low hardness of Os metal  with hexagonal close-packed
crystal structure is related to the metallic boning character.  To
improve the hardness of osmium, incorporation of small and covalent
main group elements (such as boron, carbon, nitrogen, and/or oxygen)
into osmium metal has been proposed in experimental and theoretical
studies \cite{Lundin98, Zheng05,Chumbe05,Kaner05}.  OsO$_2$ is
expected to have a covalent bonding and a high bulk modulus
\cite{Lundin98}, analogous to RuO$_2$.  Although OsC and OsN
compounds are not yet synthesized successfully,  recent \textit{ab
initio} calculations predict that \cite{Zheng05} OsC in hexagonal WC
structure has a bulk modulus of 396 GPa and is a promising superhard
material.  For one of binary compounds in Os-B system, OsB$_2$,
recent experiment  measures show that \cite{Chumbe05} OsB$_2$ is an
ultra-incompressible and hard material with bulk modulus of 365-395
GPa and hardness of $\geq$ 2000 kg/mm$^2$.

In order to investigate the mechanical properties of OsB$_2$ and
understand its high stiffness and hardness, we have evaluated the
nine independent elastic constants including bulk modulus and shear
modulus of OsB$_2$ using \textit{ab initio} pseudopotential
calculations. Our results show that the behavior of this compound
can be understood on a fundamental level in terms of their
electronic band structure. The unusual hardness originates from the
strong covalent bonding character between the $p$ bonding states of
boron atoms  and the $d$ orbitals  of osimum atom,  and the strong
covalent bonding between boron and boron atoms. The
ultra-incompressibility exhibits anisotropy ($C_{33} >C_{11} >
C_{22}$), which is due to the different bonding character of
interlayer along $c$, $a$, and $b$-directions of crystal.

\section{Method}
Because crystals deform almost in a linear elastic manner at
small strains, the increment of strain energy density of
a homogeneous and elastically deformed crystal is given by
\begin{eqnarray}
dE = \sigma_{ij}d\epsilon_{ij}\label{eq:1}
\end{eqnarray}
\noindent
where $\sigma_{ij}$ are the elements of stress tensor and $\epsilon_{ij}$ are the
elements of strain tensor. Indices $i$, $j$, $k$ and $l$ run from 1 to 3. Elastic constants are defined as
\begin{eqnarray}
c_{ijkl} \equiv \frac{\partial\sigma_{ij}}{\partial\epsilon_{ij}} = \frac{1}{V}\frac{\partial^2E}{\partial\epsilon_{ij}\partial\epsilon_{kl}}-\delta_{ij}\sigma_{kl}=C_{ijkl}-\delta_{ij}\sigma_{kl}\label{eq:2}
\end{eqnarray}
\noindent where $V$ and $\delta$ and are the volume of the unstrained crystal and Kronecker delta function, respectively. At zero hydrostatic pressure $c_{ijkl}$ = $C_{ijkl}$.
The internal energy $E(V,\{\epsilon_{ij}\})$of a
crystal under a general strain $\epsilon_{ij}$ can be expressed by
expanding the internal energy $E(V)$ of the deformed
crystal with respect to the strain tensor, as
\begin{eqnarray}
E(V,\{\epsilon_{ij}\}) = E(V,0) + V \sum_{i,j=1}^{3}\sigma_{ij}\epsilon_{ij} + \frac{V}{2}\sum_{i,j,k,l=1}^{3}C_{ijkl}\epsilon_{ij}\epsilon_{kl} + \ldots\label{eq:3}
\end{eqnarray}
\noindent where $E(V,0)$ is the corresponding ground state energy. If one uses Voigt notation ($xx\rightarrow 1$ , $yy\rightarrow 2$, $zz\rightarrow 3$, $yz\rightarrow 4$, $xz\rightarrow 5$ and $xy\rightarrow 6$) for the strain tensors subscripts (ij, kl) in the the above expression,  Eq. (\ref{eq:3}) can be rewritten as
\begin{eqnarray}\label{eq:4}
E(V,\{\epsilon_{i}\}) = E(V,0) + V \sum_{i=1}^{6}\sigma_{i}e_{i} + \frac{V}{2}\sum_{i,j=1}^{6}C_{ij}e_{i}e_{j} + \ldots
\end{eqnarray}
\noindent
with the strain tensor
\begin{eqnarray}\label{eq:5}
\epsilon=
\left(
\begin{array}{ccc}
e_1 & \frac{1}{2}e_6 & \frac{1}{2}e_5\\
\frac{1}{2}e_6 & e_2 & \frac{1}{2}e_4\\
\frac{1}{2}e_5 & \frac{1}{2}e_4 & e_3 \\
\end{array}\right)
\end{eqnarray}
\noindent Therefore, one can apply small elastic strains and
calculate the change of energy or stress to obtain elastic
constants. Direct calculation of elastic constants are possible from
first-principles total energy calculations and symmetry-general
least-squares extraction method \cite{Page02}. The independent
elastic constants for crystals can be determined by selectively
imposing strains either individually or in combination along
specific crystallographic directions~\cite{Ravindran98,Neumann99,Beckstein01,Mattesini03,Ravindran01}. Table \ref{tab:1} illustrates
the deformations employed in determining independent elastic
constants of orthorhombic crystal~\cite{Ravindran98,Beckstein01}.

First-principles total energy calculations on OsB$_2$ were performed
using a plane-wave pseudopotential method as implemented in the VASP
code \cite{GK1,GK2}. The exchange-correlation energy functional was
treated with the local density approximation (LDA) given by Ceperley
and Alder \cite{Ceperley80} as parameterized by Perdew and Zunger
\cite{Perdew81}. Electron-ion interaction was represented by the
projector augmented wave (PAW) method \cite{Blochl94,Kresse99} and
wave functions were expanded by the plane waves up to an energy
cutoff of 500 eV. Brillouin-zone integrations were approximated
using the special {\em k}-point sampling of Monhkorst-Pack scheme
\cite{HJJD} with a 9$\times$13$\times$9 grid. The total energy
convergence test showed that convergency to within 1 meV/atom was
achieved with the above calculation parameters. The bulk unit cell
lattice vectors and atomic coordinates of orthorhombic OsB$_2$ were
relaxed at a series of fixed volumes by optimizing both the forces
and stresses. The residual force and stress in the equilibrium
geometry are of the order of 0.01 eV/\AA $ $ and 10$^{-3}$ GPa,
respectively. The obtained energies were fitted with the third-order
Birch-Murnaghan equation of state (EOS) \cite{Birch47} to give the
equilibrium volume and the minimum energy. The final calculated cell
parameters are given in Table \ref{tab:2}, which are in good
agreement with the available experimental values.

\section{Results and Discussions}
\subsection{Structural properties}
The single-crystal of OsB$_2$ was found to have the orthorhombic
symmetry with the space group \textit{Pmmn} (No. 59) and lattice
constants $a$ = 4.6832 \AA, $b$ = 2.8717 \AA $ $ and $c$ = 4.0761
\AA, using the ploycrystalline $x$-ray diffraction
techniques.\cite{Roof62} In the orthorhombic unit cell of OsB$_2$,
the Os atoms were determined at the Wyckoff position of 2$a$
($\frac{1}{4}$, $\frac{1}{4}$, $z$) and 4$f$ ($u$, $\frac{1}{4}$,
$v$) for B atoms \cite{Roof62}. The crystal structure of OsB$_2$ is
illustrated in Fig.\ref{fig:1}, especially, the layer structure
repeated along $c$-(or $z$-)direction with Os-B-B-Os sandwich
structure is also shown. The orthorhombic unit cell of OsB$_2$
contains two formula unit (f.u.). The calculated lattice constant
are $a$ = 4.6444 \AA, $b$ = 2.8505 \AA $ $ $c$ = 4.0464 \AA, which
are in good agreement with the available experiment values
\cite{Roof62} and previous spin polarized LDA (namely, LSDA)
calculation results ($a$= 4.6433 \AA, $b$ = 2.8467 \AA $ $ $c$ =
4.0432 \AA) \cite{Chen05}. We should pointed out that we have also
performed the spin polarized LDA calculation on OsB$_2$ and found
that there is no difference between the results of LDA and LSDA. The
calculated volume (53.57 \AA$^{3}$/f.u.) is 2.2\% smaller than
experimental value, which is typical for the LDA approximation to
DFT. The calculated atomic coordinate parameters are $z$ = 0.1538
for Os atoms, $u$ = 0.0588 and $v$ = 0.6378 for B atoms.  We have
also calculated the equation of state (the pressure versus volume)
by taking the volume derivative of the fitted total energy obtained
from our LDA calculation, the result along the available
experimental data is shown in Fig. \ref{fig:2}. The bulk modulus and
its pressure derivative for OsB$_2$, obtained by fitting the total
energy curve to third-order Birch-Murnaghan equation of state (EOS)
\cite{Birch47}, are 336.1 GPa and 4.27 in our LDA calculations. This
bulk modulus is 7.9\% $\sim$ 15\% smaller than the experimental
value (365-395 GPa) \cite{Chumbe05}. Note that the corresponding
derivative of bulk modulus to pressure in experimental studies
\cite{Chumbe05} is fixed in the range between 4 and 1.4.  In order
to further asses the effects of the exchange-correlation
approximation, we have also studied structural properties of OsB$_2$
using the generalized gradient approximation (GGA) \cite{Perdew92}.
GGA consistently yields a larger volume than experiment, while LDA
consistently gives a smaller volume than GGA (see Table
\ref{tab:2}). Since the GGA calculation gives a smaller bulk modulus
(303.45 GPa) of OsB$_2$ than the experimental data, in the following
sections we will mainly present and discuss the results obtained by
LDA calculations.

\subsection{Elastic constants}
The nine independent elastic constants for OsB$_2$ were determined
by imposing three different strains given in Table \ref{tab:1} on
the equilibrium lattice of orthorhombic unit cell and fitting the
dependence of the resulting change in energy on the strain. Up to
seven deformations, $\gamma$ from -0.009 to 0.009 in steps of 0.003,
were applied to determine each elastic constant. The ground state
energies of OsB$_2$ unit cell were determined for all these
deformation strains. The $E(\gamma)$ curves are well fitted by the
third-order polynomials in $\gamma$ (Eq.(\ref{eq:4})), as can be
seen from the small standard errors in the calculated $C_{ij}$. The
elastic constants were then calculated using the energy relations listed in Table
\ref{tab:1}. During the calculations on all elastic constants of
orthorhombic OsB$_2$, relaxation of the coordinate of ions in the
strained lattice was carried out for achieving accurate results. The
nine independent elastic constants for orthorhombic OsB$_2$ are
listed in Table \ref{tab:3}, which experimental data on elasticity
is not available, except for bulk modulus. The nine independent
elastic constants of orthorhombic OsB$_2$ are $C_{11}$ = 597.0 GPa,
$C_{12}$ = 198.1 GPa, $C_{13}$ = 206.1 GPa, $C_{22}$ = 581.2 GPa,
$C_{23}$ = 142.6 GPa, $C_{33}$ = 825.0 GPa, $C_{44}$ = 70.1 GPa,
$C_{55}$ = 212.0 GPa, and $C_{66}$ = 201.3 GPa in our calculations.

Under the Voigt approximation~\cite{Voigt28,Ravindran98}, the elastic constants for polycrystalline aggregates are described by Voigt shear modulus ($G_{V}$) and Voigt bulk modulus ($B_{V}$). For the orthorthombic lattice, $G_{V}$ and $B_{V}$ are expressed in the following equations~\cite{Ravindran98}: 
\begin{eqnarray}
G_{V}=\frac{1}{15}(C_{11}+C_{22}+C_{33}-C_{12}-C_{13}-C_{23})+\frac{1}{5}(C_{44}+C_{55}+C_{66})\label{eq:6}
\end{eqnarray}
\begin{eqnarray}
B_{V} =
\frac{1}{9}(C_{11}+C_{22}+C_{33}+2C_{12}+2C_{13}+2C_{23})\label{eq:7}.
\end{eqnarray}
The calculated Voigt shear constant of OsB$_2$ is about 193.8 GPa. The Voigt bulk modulus calculated
from the theoretical values of the elastic constants  is
344.1 GPa. It agrees well with the one extracted from the fit to the
third-order Birch-Murnaghan equation of state, 336.1 GPa. Since the Young's modulus ($E$) and Poisson's ratio ($v$) for an isotropic materials can be obtained from the following relations~\cite{Grimvall99,Ravindran98}:
 \begin{eqnarray}
E=\frac{9BG}{3B+G}\label{eq:8}
\end{eqnarray}
\begin{eqnarray}
v=\frac{3B-2G}{2(3B+G)}\label{eq:9}.
\end{eqnarray}
The calculated Young's modulus  and Poisson's ratio of OsB$_2$ are 489.5 GPa and 0.263, respectively. 

The requirement of mechanical stability in a orthorhombic crystal
leads to the following restrictions on the elastic
constants:\cite{Grimvall99,Beckstein01}
\begin{eqnarray}\label{eq:13}
\nonumber
(C_{11}+C_{22}-2C_{12}) > 0,  \\
\nonumber
(C_{11}+C_{33}-2C_{13}) > 0,  \\
\nonumber
(C_{22}+C_{33}-2C_{23}) > 0,  \\
C_{11} > 0, C_{22} >0, C_{33} >0, \\
\nonumber
C_{44} > 0, C_{55} >0, C_{66} >0, \\
\nonumber
(C_{11}+C_{22}+C_{33}+2C_{12}+2C_{13}+2C_{23}) > 0 .
\end{eqnarray}
\noindent Since $B_{0}$ is a weighted average of $C_{ii}$ and $C_{ij}$ (indices $i$ and $j$ running from 1 to 3), these conditions also lead to a restriction on the magnitude of $B_{0}$. Namely, $B_{0}$ is required to be intermediate in value between $\frac{1}{3}$($C_{12} + C_{13} + C_{23}$) and $\frac{1}{3}$($C_{11} + C_{22} + C_{33}$)~\cite{Grimvall99,Beckstein01},
\begin{eqnarray}\label{eq:14}
\frac{1}{3}(C_{12} + C_{13} + C_{23}) < B_{0} < \frac{1}{3}(C_{11} + C_{22} + C_{33})
\end{eqnarray}
\noindent The elastic constants of orthorhombic OsB$_2$ in Table
\ref{tab:3} obey these stability conditions (Eq.(\ref{eq:13}) and
(\ref{eq:14})). In particular, $C_{12}$ is smaller than the average
of $C_{11}$ and $C_{22}$, $C_{13}$ is smaller than the average of of
$C_{11}$ and $C_{33}$, $C_{23}$ is smaller than the average of
$C_{22}$ and $C_{33}$, and the bulk modulus is smaller than the
average of $C_{11}$, $C_{22}$, and $C_{33}$ but larger than the
average of $C_{12}$, $C_{13}$, and $C_{23}$.

\subsection{Electronic structures and analysis of chemical bonding}
The energy band structure and density of states (DOS) (including
total, site and angular momentum decomposed DOS) of orthorhombic
OsB$_2$ are shown in Fig. \ref{fig:2} and \ref{fig:3}, respectively.
It evidently exhibits a metallic character. The valence bands in
Fig. \ref{fig:2} have a width of about 15 eV, and are split into two
disjointed groups, respectively. The lower group has a width of
about 5 eV and the upper one up to the Fermi level ($E_{F}$) has a
width of about 10 eV. Combining from the DOS figures (\ref{fig:3}),
the lower group in the energy range between -15 eV and  -10 eV below
the Fermi level ($E_{F}$) is largely contributed by B-2$s$ levels
and small amount of Os-6$s$, the upper one in the energy range from
-12 eV to $E_{F}$ is mainly composed by Os-5$d$ orbitals and B-2$p$.
It  suggests that the strong hybridization between B-2$p$ and
Os-5$d$ orbitals occur below the Fermi level. Additionally, $d$
orbitals of Os atom are contributed most to the total DOS at
$E_{F}$. These findings are consistent with the previous studies on electronic structure of OsB$_2$~\cite{Ivanovskii99,Chen05}.

The large values and anisotropy of elastic constants of OsB$_2$ can
be understood by examining the nature of chemical bonding between
the atoms. In order to understand the bonding between the Os-B,
Os-Os, and B-B atoms, the lines charge  density along nearest
neighbor Os$-$B, Os$-$Os, B$-$B atoms are illustrated in Fig.
\ref{fig:5}. It is clearly seen that charge highly accumulates
around Os atom, and Os-B, Os-Os and B-B bondings exhibit covalent,
metallic, covalent characters, respectively. It also supported by
above discussion that the $d$ states of Os atoms and $p$ states of B atoms have
a strong hybridization seen in density of states.  In order to
further reveal the topology of Os-Os, Os-B and the B-B bondings, we
illustrate the contour plots of electron density in different
crystallographic planes. Fig. \ref{fig:6} illustrates the
two-dimensional electron charge density distribution in the planes
with normal vector along $<$100$>$, $<$010$>$ and $<$001$>$,
respectively. From Fig. \ref{fig:6}(a) and (d), in which planes with
the normal vector along $<$100$>$ and central point at Os1 atom
($\frac{1}{4}$, $\frac{1}{4}$, $z$), and with the normal vector
along $<$001$>$ and central point at Os1 atom, respectively,  it can
be clearly seen that there is no strong bonding between Os$-$Os
atoms as there is no accumulation of charges in excess of the
background charge density of the plane. It appears that the Os$-$Os
bonding is predominantly metallic in OsB$_2$.  Fig. \ref{fig:6}(e)
and (f), in which planes with the normal vector along $<$001$>$ and
central point at B1 atom ($u$, $\frac{1}{4}$, $v$), and through B1,
B2, B3 and B4 atoms, respectively,  It also can be clearly seen that
there is strong covalent bonding between B-B atoms. The nature of
Os-B and B-B bonding is illustrated in Fig. \ref{fig:6}(c) of the
two-dimensional contour electron density map with normal vector
along $<$010$>$ and central point at  Os1 atom ($\frac{1}{4}$,
$\frac{1}{4}$, $z$). It can be seen that there is a  charge
accumulation of electron charge between the Os-B atoms, which is
consistent with the idea of B-2p and Os-5d hybridization noted in
Fig. \ref{fig:4}. This also clearly confirms both  B-B  and Os-B
bondings are covalent character.  Because  the orthorhombic OsB$_2$
is stacked by Os-B-B-Os atomic layer structures in the $c$ (or $z$)
direction of crystal, these mean that the atom bondings of
interlayer are strong covalent characters of Os-B and B-B. This also
explains the high value of elastic constant $C_{33}$ (825.0 GPa) and
a relatively lower value of elastic constant $C_{11}$ (597.0 GPa)
and $C_{22}$ (581.2 GPa).

\section{Conclusions}

The structural parameters, elastic constants, and electronic
structures of OsB$_2$ have been calculated and analyzed. Our
calculated structural parameters including the lattice constants and
internal parameters are in good agreement with the experiment data.
The LDA calculations yielded the bulk modulus (336.1 GPa) of OsB$_2$
,which is more close to the experimental data than that of GGA
calculations (303.45 GPa). The LDA calculations also predicted that
the independent elastic constants of OsB$_2$ are  $C_{11}$ = 597.0
GPa, $C_{12}$ = 198.1 GPa, $C_{13}$ = 206.1 GPa, $C_{22}$ = 581.2
GPa, $C_{23}$ = 142.6 GPa, $C_{33}$ = 825.0 GPa, $C_{44}$ = 70.1
GPa, $C_{55}$ = 212.0 GPa, and $C_{66}$ = 201.3 GPa, respectively.
The incompressibility of OsB$_2$ exhibits anisotropy ($C_{33} >
C_{11}
>C_{22}$) and the $c$-direction of crystal  is the least
compressible  due to the difference of atom bonding character of
interlayer along different crystalline planes. The detailed study of
the electronic structure and charge density redistribution reveals
the strong covalent Os-B atoms bonding and B-B atoms bonding play a
important role in the incompressibility and hardness of
OsB$_2$.
\section{Acknowledgements}
The author acknowledge support from Shanghai Postdoctoral Science
Foundation under Grant No. 05R214106. The computation was performed
at  Supercomputer Center of Fudan. The XCRYSDEN package
\cite{Kokalj99}  and  LEV00 code \cite{Kantorovich02} were used for
generation of ball-and-stick model and calculation of charge density
maps, respectively.


\clearpage
\newpage
\begin{table*}
\caption{\label{tab:1} Parametrization of the strains used to calculate the
elastic constants of orthorhombic  OsB$_2$. The energy expressions reported
on the third column refer to Eq.~\ref{eq:4}.}
\begin{ruledtabular}
\begin{tabular}{ccc}
Strain  &Parameters (unlisted $e_i$ = 0) & $\Delta E/V$ in $O$($\gamma^2$) \\
\hline
$\epsilon^1$ & $e_1=\gamma$   & $\frac{1}{2}C_{11}\gamma^2$ \\
$\epsilon^2$ & $e_2=\gamma$   & $\frac{1}{2}C_{22}\gamma^2$ \\
$\epsilon^3$ & $e_3=\gamma$   & $\frac{1}{2}C_{33}\gamma^2$ \\
$\epsilon^4$ & $e_1=2\gamma$, $e_2=-\gamma$, $e_3=-\gamma$ & $\frac{1}{2}$(4$C_{11}$ - 4$C_{12}$ - 4$C_{13}$ + $C_{22}$ + 2$C_{23}$ + $C_{33}$)$\gamma^2$ \\
$\epsilon^5$ & $e_1=-\gamma$, $e_2=2\gamma$, $e_3=-\gamma$ & $\frac{1}{2}$($C_{11}$ - 4$C_{12}$ + 2$C_{13}$ + 4$C_{22}$ - 4$C_{23}$ + $C_{33}$)$\gamma^2$ \\
$\epsilon^6$ & $e_1=-\gamma$, $e_2=-\gamma$, $e_3=2\gamma$ & $\frac{1}{2}$($C_{11}$ + 2$C_{12}$ - 4$C_{13}$ + $C_{22}$ - 4$C_{23}$ + 4$C_{33}$)$\gamma^2$ \\
$\epsilon^7$ & $e_4=\gamma$   & $\frac{1}{2}C_{44}\gamma^2$ \\
$\epsilon^8$ & $e_5=\gamma$   & $\frac{1}{2}C_{55}\gamma^2$ \\
$\epsilon^9$ & $e_6=\gamma$   & $\frac{1}{2}C_{66}\gamma^2$ \\
\end{tabular}
\end{ruledtabular}
\end{table*}

\begin{table*}
\caption{\label{tab:2}Calculated lattice parameters ($a$, $b$ and $c$, in \AA), volumes of unit cell ($V$, in \AA$^3$), and bulk modulus ($B$, in GPa) of orthorhombic OsB$_2$, along a comparison with other theoretical work and available experimental data.}
\begin{ruledtabular}
\begin{tabular}{c cc cc c}
Property&\multicolumn{2}{c}{This work}&\multicolumn{2}{c}{Previous\footnotemark[1]}  & Experiment \\
\hline
     &  LDA    &    GGA   &   LSDA    &  LSDA+SO  &            \\
$a$  & 4.6444  &  4.7049  &   4.6433  &  4.6383   & 4.6832\footnotemark[2]   \\
$b$  & 2.8505  &  2.8946  &   2.8467  &  2.8437   & 2.8717\footnotemark[2] \\
$c$  & 4.0464  &  4.0955  &   4.0432  &  4.0388   &  4.0761\footnotemark[2]    \\
$V$  & 53.57   &  55.78   &   53.44   &  53.27    &  54.82\footnotemark[2]    \\
$B$  & 336.1  &  303.45 & 364.7   &  385.4  & 365-395\footnotemark[3]   \\
\end{tabular}
\end{ruledtabular}
\footnotetext[1]{see Ref.\onlinecite{Chen05}}
\footnotetext[2]{see Ref.\onlinecite{Roof62}}
\footnotetext[3]{see Ref.\onlinecite{Chumbe05}}
\end{table*}

\begin{table*}
\caption{\label{tab:3} Calculated elastic constants of orthorhombic OsB$_2$. Note that a
relaxation of the atomic positions was carried out.  All values are in units of GPa.}
\begin{ruledtabular}
\begin{tabular}{ccc ccc ccc}
 $C_{11}$ & $C_{12}$ & $C_{13}$ & $C_{22}$ & $C_{23}$ & $C_{33}$ & $C_{44}$ & $C_{55}$ & $C_{66}$  \\
\hline
597.0 &   198.1  & 206.1 &  581.2 &  142.6 &  825.0  &  70.1 &  212.0 &  201.3 \\
\end{tabular}
\end{ruledtabular}
\end{table*}

\begin{figure}
\includegraphics*[scale=0.8]{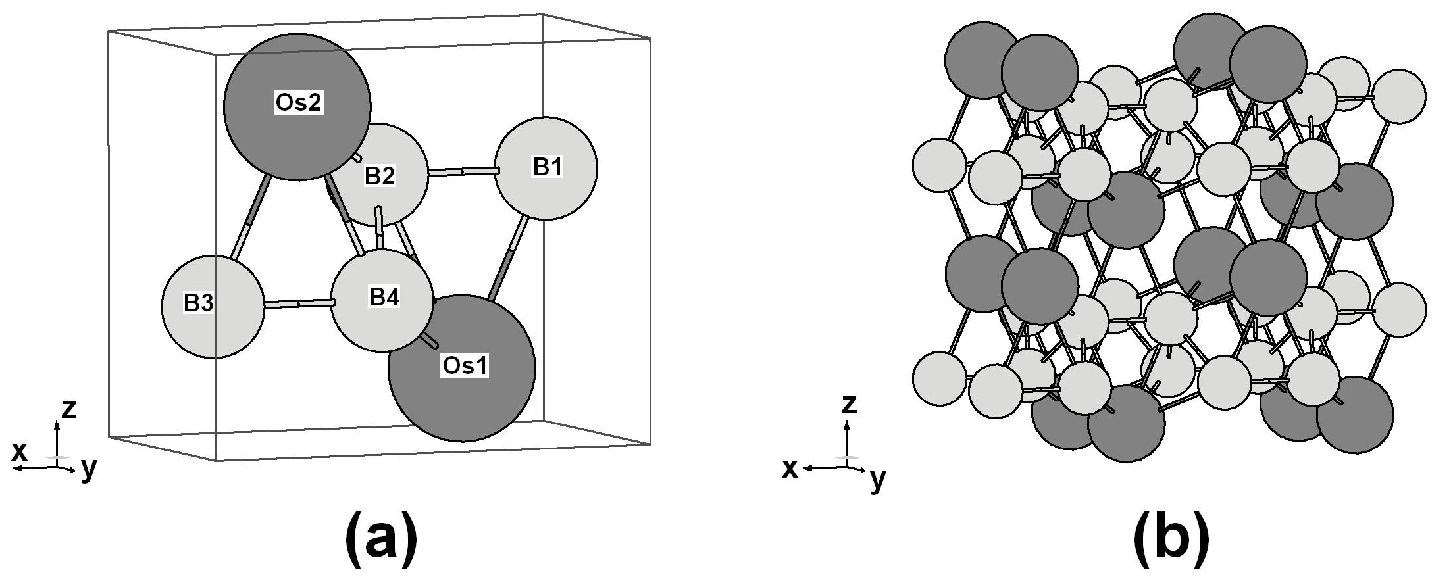}
\caption{\label{fig:1} }
\end{figure}

\begin{figure}
\includegraphics*[scale=0.9]{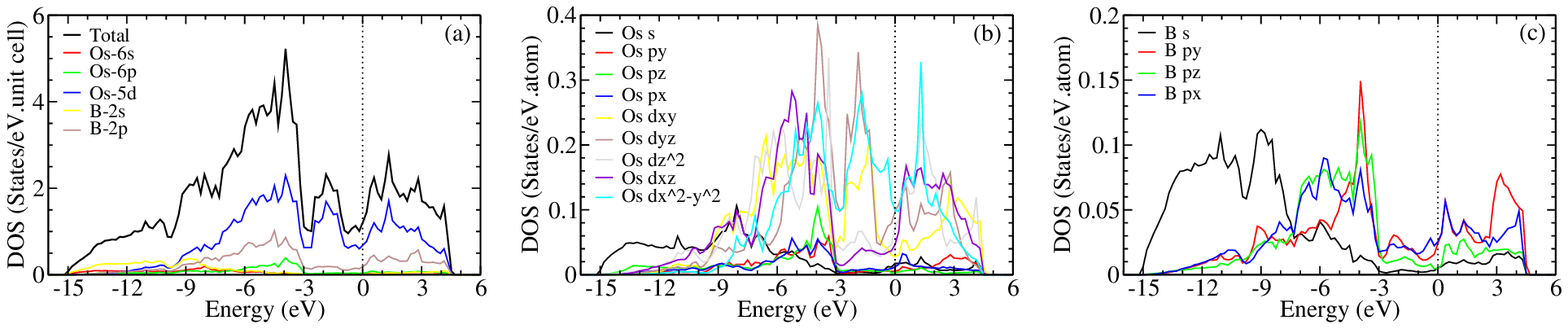}
\caption{\label{fig:4}}
\end{figure}

\begin{figure}
\includegraphics*[scale=0.9]{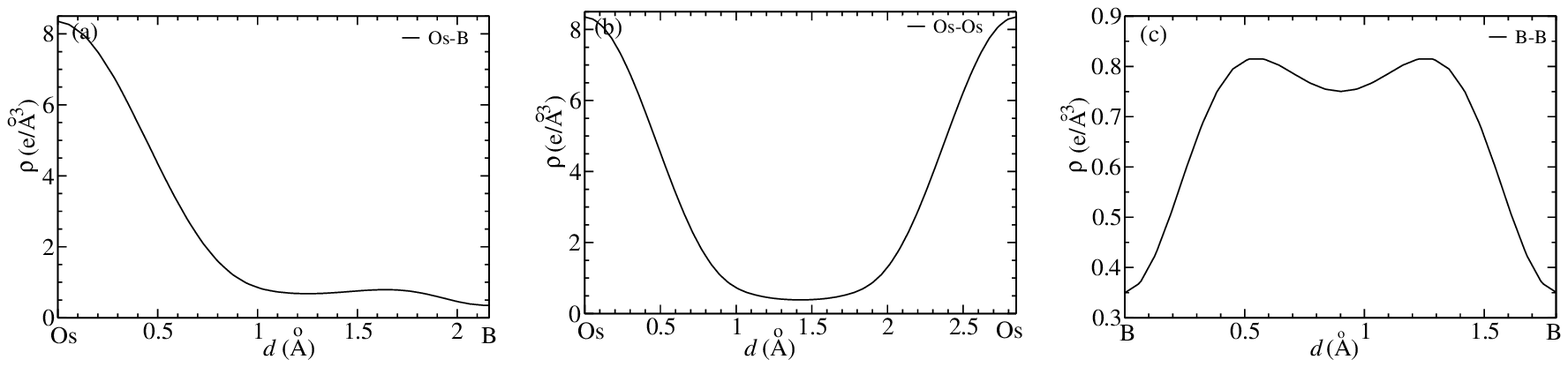}
\caption{\label{fig:5} }
\end{figure}

\begin{figure}
\includegraphics*[scale=0.9]{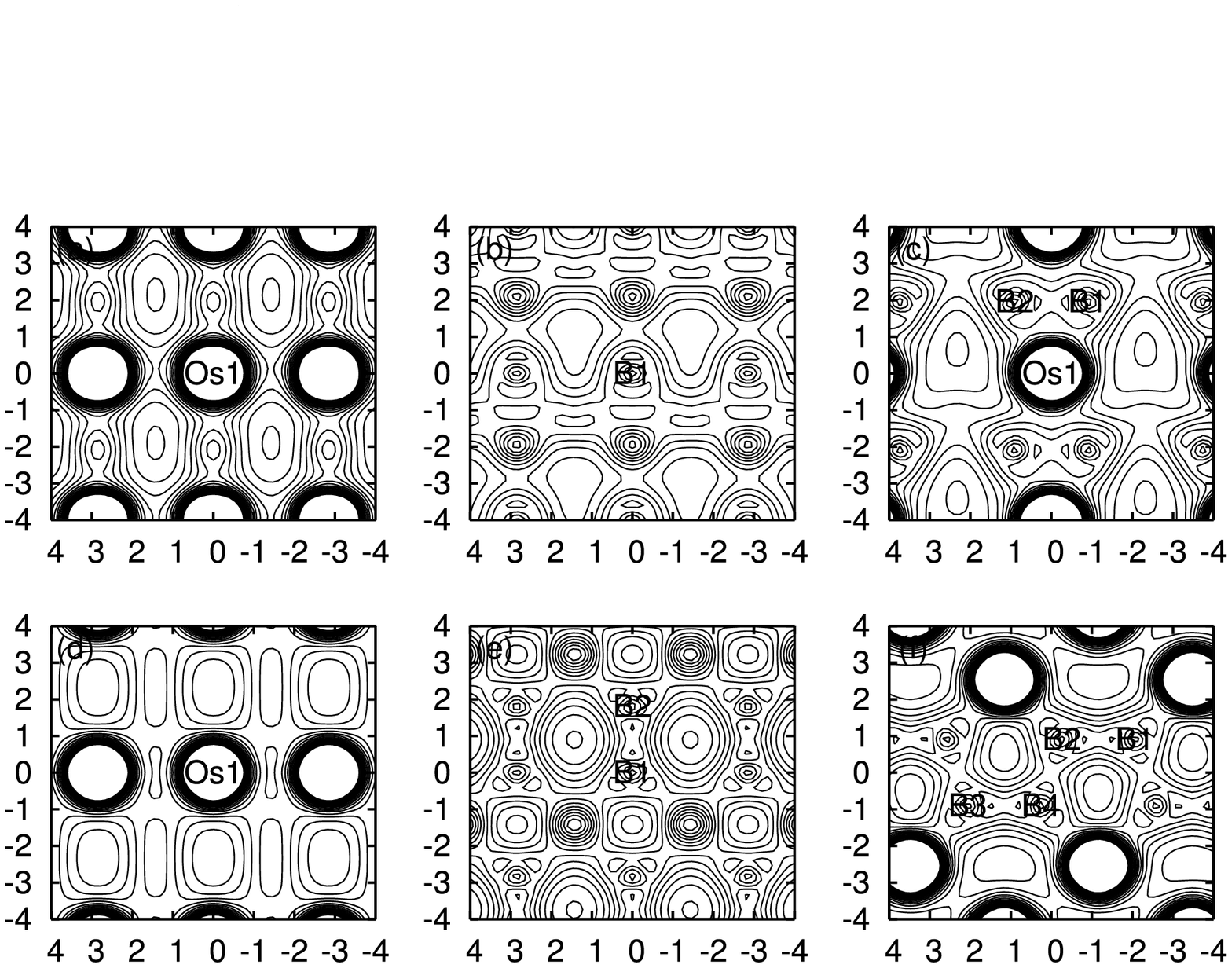}
\caption{\label{fig:6} }
\end{figure}

\clearpage
\newpage
Figure Captions
\begin{itemize}
\item FIG \ref{fig:1}: The ball-and-stick model for structure of
orthorhombic OsB$_2$. (a) The unit cell containing a single formula
unit. Two Os atoms at Wyckoff positions of 2$a$: Os1 ($\frac{1}{4}$,
$\frac{1}{4}$, $z$) and Os2 ($\frac{3}{4}$, $\frac{3}{4}$, $-z$).
Four B atoms at Wyckoff positions of 4$f$: B1 ($u$, $\frac{1}{4}$,
$v$), B2 ($-u+\frac{1}{2}$, $\frac{1}{4}$, $v$), B3 ($-u$,
$\frac{3}{4}$, $-v$), and B4 ($u+\frac{1}{2}$, $\frac{3}{4}$, $-v$).
(b) The supercell repeated in three dimensions by
2$\times$2$\times$2, in order to illustate the layers in $z$(or
$c$)-direction. Os atoms are shown as dark gray balls and boron
atoms as light gray balls.

\item FIG \ref{fig:2}:  Calculated pressure versus fractional unit
cell volume for OsB$_2$, along a comparison with available
experiment data.

\item FIG \ref{fig:3}: Electronic band structure along the
high-symmetry direction of the Brillouin zone for OsB$_2$.
High-symmetry points are labeled as the following~\cite{specialk}: $\Gamma$ = (0.0,
0.0, 0.0), Y = (0.0, 0.5, 0.0), S = (0.5, 0.5, 0.0), X = (0.5, 0.0,
0.0), and Z = (0.0, 0.0, 0.5). The zero of energy is set as the
Fermi level and shown in dotted line.

\item FIG \ref{fig:4}: (Color online) Total, site and angular
momentum-decomposed density of states (DOS) of OsB$_2$.

\item FIG \ref{fig:5}: Charge density at the equilibrium lattice
parameters of OsB$_2$, along the lines between (a) Os and B atoms,
(b) Os and Os atoms, and (c) B and B atoms.

\item FIG \ref{fig:6}: Contours of the charge density at the
equilibrium lattice parameters of OsB$_2$, in the planes (a) with
normal vector along $<$100$>$ and central point at Os1 atom, (b)
with normal vector along $<$100$>$ and central point at B1 atom, (c)
with normal vector along $<$010$>$ and central point at Os1 atom,
(d) with normal vector along $<$001$>$ and central point at Os1
atom, (e) with normal vector along $<$001$>$ and central point at B1
atom, and (f) through B1, B2, B3 and B4 atoms. Charge density is in
an increment of 0.1 \textit{e}/\AA$^3$ from 0 to 2.1
\textit{e}/\AA$^3$ and all distances are in \AA.

\end{itemize}

\clearpage
\newpage

\clearpage
\newpage

\begin{figure}
\includegraphics*[scale=0.6]{eos.eps}
\caption{\label{fig:2} }
\end{figure}

\clearpage
\newpage

\begin{figure}
\includegraphics*[scale=0.5]{bnd.eps}
\caption{\label{fig:3} }
\end{figure}

\clearpage
\newpage

\clearpage
\newpage

\clearpage
\newpage

\end{document}